\documentclass[conference]{IEEEtran}
\IEEEoverridecommandlockouts
\usepackage{cite}
\usepackage{amsmath,amssymb,amsfonts}
\usepackage{algorithmic}
\usepackage{algorithm}
\usepackage{graphicx}
\usepackage{subfig}
\usepackage{textcomp}
\usepackage{xcolor}
\usepackage{url}

\usepackage{epstopdf}
\def\BibTeX{{\rm B\kern-.05em{\sc i\kern-.025em b}\kern-.08em
    T\kern-.1667em\lower.7ex\hbox{E}\kern-.125emX}}

\begin{document}

\title{WDMoE: Wireless Distributed Large Language Models with Mixture of Experts\\
}
\author{Nan Xue, Yaping Sun, Zhiyong Chen, Meixia Tao, Xiaodong Xu, Liang Qian, Shuguang Cui, Ping Zhang\\
\thanks{
    N. Xue, Z. Chen, M. Tao and L. Qian are with the Cooperative Medianet Innovation
Center, Shanghai Jiao Tong University, Shanghai 200240, China. (email:
\{nan.xue, zhiyongchen, mxtao, lqian\}@sjtu.edu.cn)

Y. Sun is with the Department of Broadband Communication, Pengcheng
Laboratory, Shenzhen 518000, China. Y. Sun is also
with the Future Network of Intelligent Institute (FNii), the Chinese University
of Hong Kong (Shenzhen), Shenzhen 518172, China (email: sunyp@pcl.ac.cn).

X. Xu and P. Zhang are with the Beijing University of Posts and Telecommunications, Beijing 100876, China, and affiliated with the Department of
Broadband Communication, Peng Cheng Laboratory, Shenzhen 518000, China
(email: xuxd@pcl.ac.cn; pzhang@bupt.edu.cn).

S. Cui is with the School of Science and Engineering (SSE) and the
Future Network of Intelligent Institute (FNii), the Chinese University of Hong
Kong (Shenzhen), Shenzhen 518172, China. S. Cui is also affiliated with the
Department of Broadband Communication, Peng Cheng Laboratory, Shenzhen
518000, China (email: shuguangcui@cuhk.edu.cn).
}
}

\maketitle

\begin{abstract}
Large Language Models (LLMs) have achieved significant success in various natural language processing tasks, but how wireless communications can support LLMs has not been extensively studied. In this paper, we propose a wireless distributed LLMs paradigm based on Mixture of Experts (MoE), named WDMoE, deploying LLMs collaboratively across edge servers of base station (BS) and mobile devices in the wireless communications system. Specifically, we decompose the MoE layer in LLMs by deploying the gating network and the preceding neural network layer at BS, while distributing the expert networks across the devices. This arrangement leverages the parallel capabilities of expert networks on distributed devices. Moreover, to overcome the instability of wireless communications, we design an expert selection policy by taking into account both the performance of the model and the end-to-end latency, which includes both transmission delay and inference delay. Evaluations conducted across various LLMs and multiple datasets demonstrate that WDMoE not only outperforms existing models, such as Llama 2 with 70 billion parameters, but also significantly reduces end-to-end latency.
\end{abstract}


\section{Introduction}
The development of large language models (LLMs), exemplified by ChatGPT\cite{openai2022chatgpt}, has garnered significant attention in recent years from academia and industry. Powerful LLMs have illuminated the concept of artificial general intelligence and demonstrated emergent capabilities. In the realm of 6G technologies, such as the metaverse, which is enhanced by projects like Sora\cite{openai2024sora}, autonomous driving systems supported by LLMs\cite{fu2024drive}, and intelligent networking solutions like those developed by Net Master\cite{huawei2024net},  large models are increasingly critical. These models are pivotal not only in advancing technological frontiers but also in creating a wide array of applications. 

This inevitably raises a question for wireless communications systems: \textbf{How can wireless communications enable LLMs?} The answer lies in achieving seamless intelligence. There are currently approaches to this: utilizing cloud-based LLMs, and  device-based LLMs\cite{li2024transformer}. Most major LLMs are deployed on cloud computing clusters. Another burgeoning area of research is device-based LLMs. Researchers are actively compressing these large models to sizes manageable for mobile devices, primarily through model quantization and distillation techniques.  

However, these approaches still face significant challenges. Cloud-based LLMs meet severe privacy issues, and users are heavily dependent on their connections to cloud servers. Device-based LLMs are constrained by limited computing capabilities, memory size and energy consumption. For example, mobile devices can generate up to 20 tokens per second on the advanced mobile AI computing platform Snapdragon 8 Gen 3, under conditions optimized jointly with Llama2-7B\cite{qualcomm2023snap}. The scaling law\cite{kaplan2020scaling} suggests that larger models tend to perform better, implying that smaller device-based LLMs may struggle to match the capabilities of their larger counterparts.

In this paper, we try to design a wireless distributed LLMs paradigm, deploying LLMs collaboratively across edge servers of base station (BS) and mobile devices, to address the aforementioned challenges. Compared to cloud deployments, edge servers are geographically closer to users, which reduce latency. Furthermore, distributing model structure across multiple locations enhances data security, making it difficult for any single location to access original user data. In addition, the wireless distributed LLMs paradigm can utilize large models with more parameters  and improve the speed of data processing compared to device-based LLMs.

\begin{figure*}[t]
    \centering
    \subfloat[]{\includegraphics[scale=0.37]{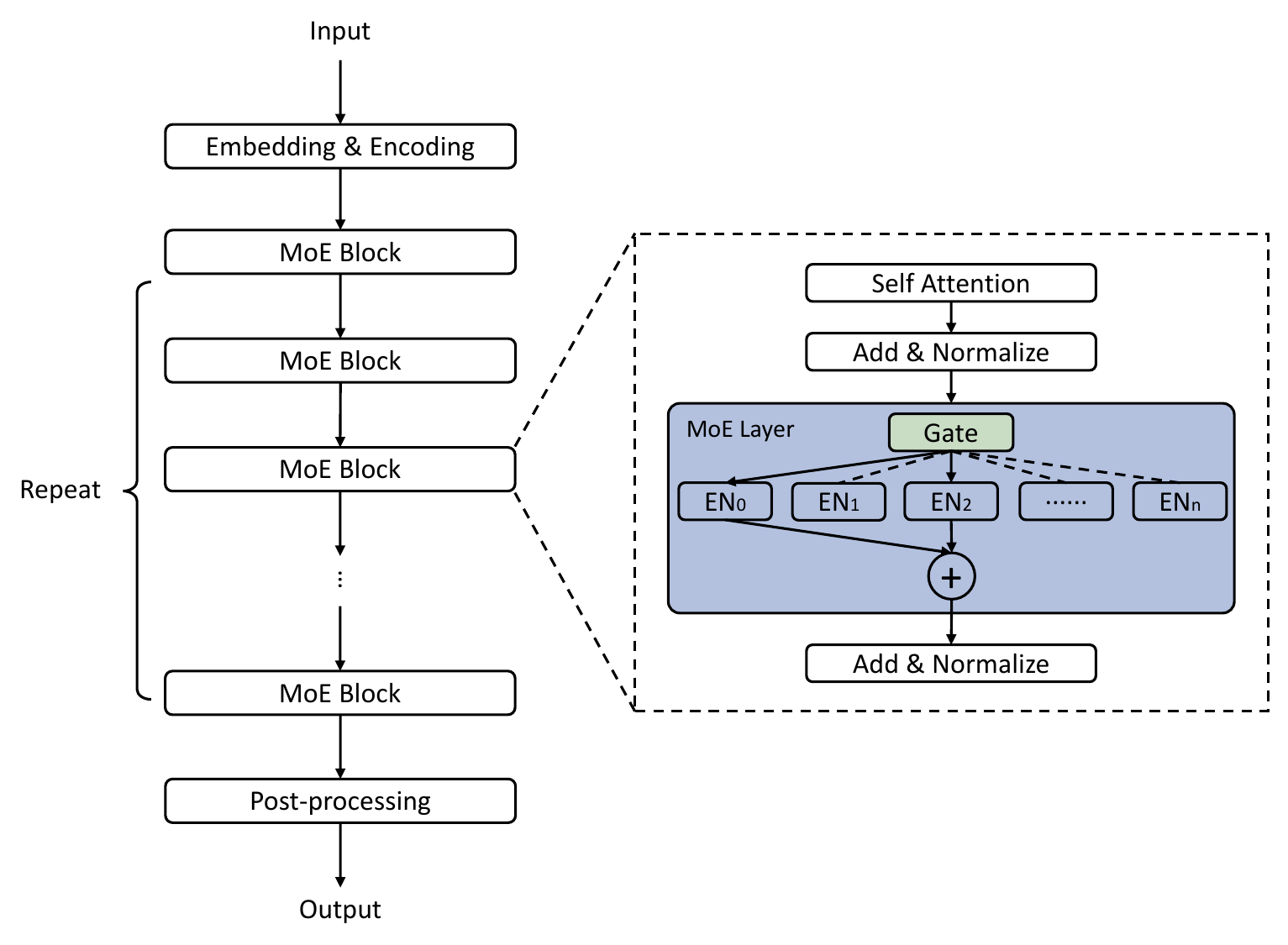}\label{moe}}
    \subfloat[]{\includegraphics[scale=0.37]{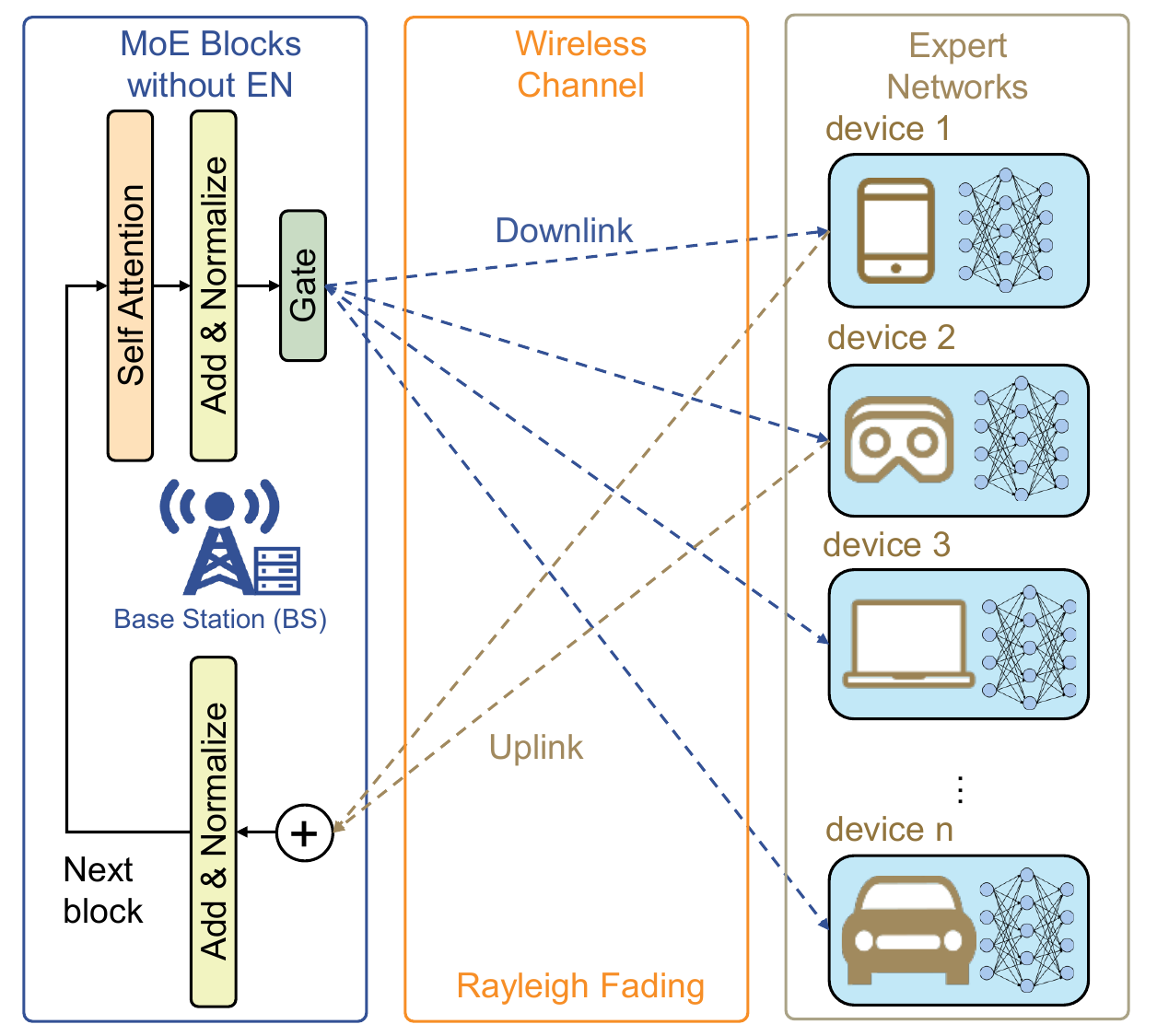}\label{sys}}\\
    \caption{(a) MoE-based LLM architecture; (b) The proposed WDMoE system model.}
    \label{mas}
\end{figure*}

To efficiently distribute LLMs across edge servers and devices, we employ the Mixture of Experts (MoE) as the model design strategy, which allows for scalable and dynamic allocation of computational and communications resources. MoE is a neural network model that extends Transformer architecture by replacing its feed-forward network (FFN) layer with a gating network and multiple expert networks \cite{lepikhin2020gshard}. This configuration introduces sparsity, enabling the model to scale up in size without a corresponding increase in computational burden. As shown in Fig.\ref{mas}\subref{moe}, the parallel relationship among experts in MoE facilitates natural robustness during the inference phase in the wireless network. Existing research \cite{vepakomma2018split,thapa2022splitfed,wu2023split} primarily focuses on the training phase, and the inherent characteristics of MoE are often overlooked \cite{hu2022distributed}. Both \cite{zhou2022mixture} and \cite{zeng2024turn} devise efficient routers that achieve either speed up or performance improvement. \cite{yi2023edgemoe} develops compression techniques for expert weights and optimizes in-memory expert management by improving the compute-I/O pipeline.

To this end, we propose the wireless distributed LLMs paradigm based on MoE, named WDMoE. This paradigm includes an expert network deployment strategy, along with a scheme for dynamically adjusting expert selection according to wireless channel conditions, without the need for additional training. Specifically, we decompose the MoE layer in LLMs by placing both the gating network and the neural network layer preceding it at BS, while the expert networks are distributed across the devices. Therefore, by leveraging the parallel characteristics of the expert networks, the proposed framework avoids scenarios in which deploying multiple expert networks on a single device could lead to processing halts due to disconnections. Meanwhile, it fully utilizes the computational resources of various devices. In LLMs based on MoE, the gating network is trained to optimize processing for each token. However, it cannot guarantee that each device maintains robust communications with BS in the wireless communications system. Therefore, we design an expert selection policy by taking into account both the optimal performance of the model and the end-to-end latency. Evaluations conducted across various LLMs and multiple datasets demonstrate that the proposed WDMoE not only outperforms existing models, such as Llama 2 with 70 billion parameters, but also significantly reduces end-to-end latency, including both wireless transmission delay and inference delay.
\section{The WDMoE Framework}
\subsection{Gating Network of MoE-based LLMs}
The network structure of MoE-based LLMs is depicted in Fig.\ref{mas}\subref{moe}, where each MoE block is based on a Transformer. In these blocks, the FFN module is replaced by an MoE layer\cite{jiang2024mixtral}. An MoE layer consists of a gating network, also known as a router, and multiple expert networks, each of which can be any type of neural network. The gating network, which is a compact neural network, processes the input token to produce a weight vector for each expert.

The number of expert networks is denoted as $n$. Let $\mathbf{x} \in \mathbf{R}^{r \times m}$ be the input token set of the gating network, where $r$ is the number of input tokens of one prompt, and $m$ is the dimension of a token's embedding. The element $x^{i}_{j} \in \mathbf{x}$ denotes the $j$-th input token of the $i$-th MoE block. This token $x^i_j$ serves as the input to the gating network, which then outputs weights $\mathbf{w}^{i}_{j}$ for the $j$-th token in the $i$-th block. Here, $\mathbf{w}^{i}_{j}$ is an $n$-dimension column vector representing weights allocated to $n$ experts. $x^i_j$ is also fed into the expert networks, whose outputs $\mathbf{y}^i_j$ is a $n$-dimension column vector. The output of an MoE layer is the weighted sum of each expert network's output as follows:
\begin{equation}
    o^i_j = {\mathbf{w}^i_j}^\intercal \mathbf{y}^i_j,
\end{equation}
where $o^i_j$ is the output of the $j$-th token through the $i$-th MoE layer. 

\subsection{Distributed Deployment of WDMoE}
Experts operate in parallel and have no impact on other experts, making this approach suitable for deployment in distributed mobile edge networks. We consider a BS with edge servers that possess powerful computing capability, enabling it to process multiple data streams simultaneously for $n$ mobile devices, as shown in Fig.\ref{mas}\subref{sys}. In the WDMoE framework, the attention mechanism and gating network are located at BS, while  only expert networks are assigned to mobile devices. Typically, the expert network consists of a simple multilayer perceptron (MLP). When a user sends a prompt, its embedding operation can be completed either locally or at BS, depending on the choice of the user. In this paper, we focus on the communications and computing costs incurred during interactions between BS and mobile devices following the first embedding module. The proposed framework can be integrated with existing splitting methods.

\subsection{Communication Model}
Let $\mathbf{U}$ denote the device set. For the $q$-th device $q \in \mathbf{U}$, the bandwidth allocated by BS is denoted by $B_{q}$. The transmission rate for the $q$-th device is formulated as:
\begin{equation}
    R_{q}^d = B_{q} \log_2 (1 + \frac{P^d_{q}g^2_{BS,q}}{N_0 B_{q}}),
    \label{comm_d}
\end{equation}
for the downlink transmission. Similarly, for the uplink tranmission, we have:
\begin{equation}
    R_{q}^u = B_{q} \log_2 (1 + \frac{P^u_{q}g^2_{q,BS}}{N_0 B_{q}}),
    \label{comm_u}
\end{equation}
where $g_{BS,q}$ and $g_{q,BS}$ represent the channel gain of downlink and uplink between BS and the $q$-th device respectively, and $N_0$ denotes the noise power spectral density. $P_{q}^d$ and $P_{q}^{u}$ denote the transmission power of downlink and uplink respectively.

\subsection{Computing Model}
In this paper, we assume that each device is equipped with at least one graphic processing unit (GPU). The expert network in this paper is an FFN. The number of floating point operations required by our expert network is:
\begin{equation}
    L^{comp} = 4 m \times m_{h} + 2 m_{h} \times m + \eta \times m_{h} + m_{h},
\end{equation}
where $m_h$ is the hidden dimension of the FFN and $\eta$ denotes the FLOPs required by the activation function in the network. Note that we assume the computing latency at BS can be ignored.

\subsection{End to End Inference Latency}
Data size is calculated by:
\begin{equation}
    L^{comm} = \epsilon \times m,
\end{equation}
where $\epsilon$ is a coefficient determined by the quantization precision.

The total delay of the $j$-th token in the $i$-th block processed by the expert $q$ is given by:
\begin{equation}
    t^i_{j, q} = \frac{L^{comm}}{R^{d}_{q}} + \frac{L^{comp}}{C_{q}} + \frac{L^{comm}}{R^{u}_{q}}, 
    \label{userl}
\end{equation}
where $C_{q}$ is the FLOPS of the GPU on the $q$-th device.

The number of tokens in a prompt can vary widely, ranging from a few to several hundreds or even thousands. In practical applications, WDMoE distributes tokens among multiple expert networks. The maximum end-to-end latency across these experts becomes a more appropriate measure. The overall prompt latency results from the accumulation of latencies as all tokens pass through all blocks. We thus have the end-to-end latency as follows:
\begin{equation}
    t^i_j = \max_{q\in\{1, 2, \cdots, k\}} \{t^i_{j, q}\},
\end{equation}
\begin{equation}
    t^{prompt} = \sum_j^p \sum_i^b t^i_j,
\end{equation}
where $b$ is the number of blocks in the model, and $p$ is the number of tokens in a prompt. $b$ and $p$ depend on the specified model and user input.

\section{Expert Selection Policy}
\begin{figure*}[t]
    \centerline{\includegraphics[scale=0.45]{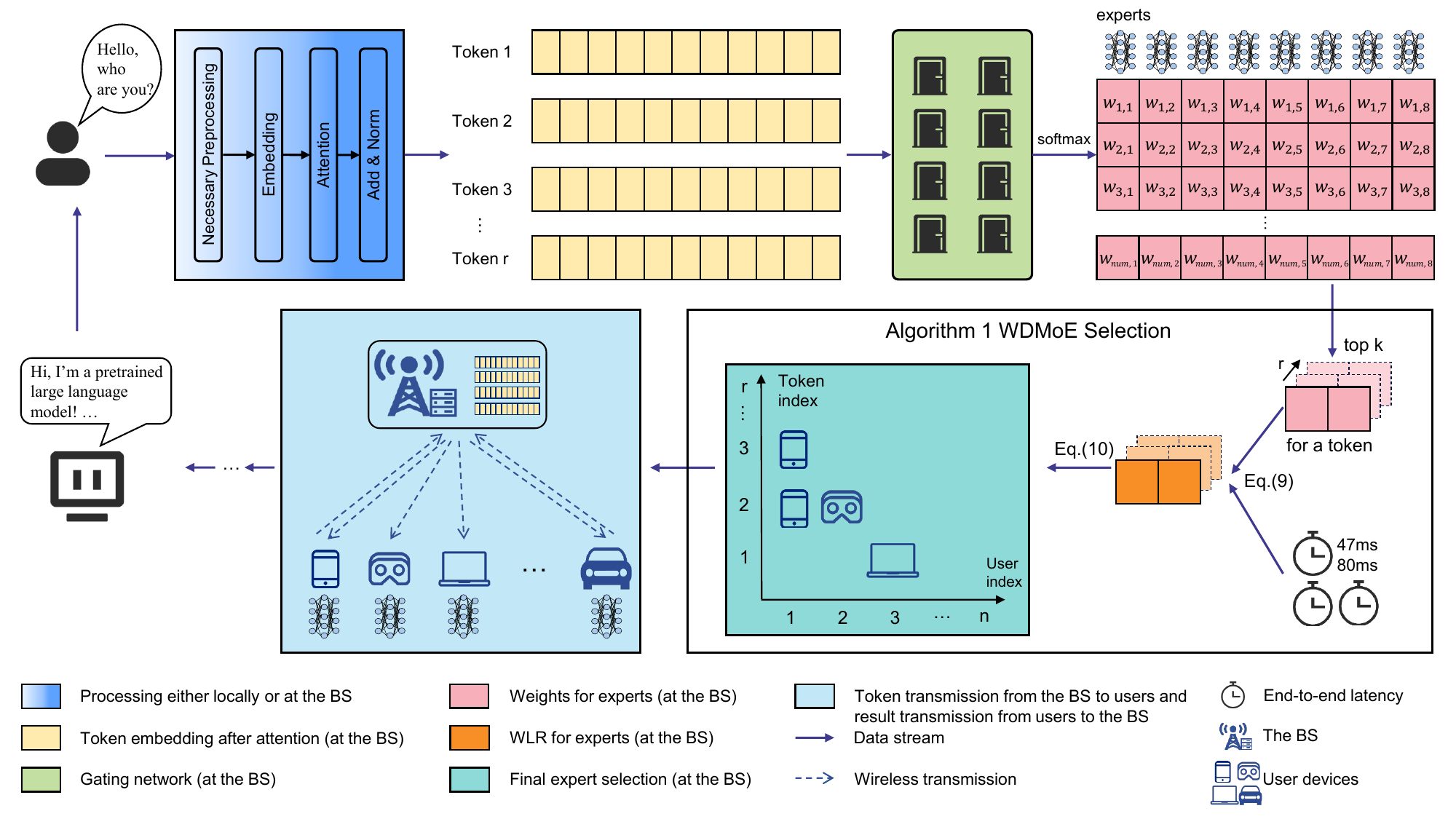}}
    \caption{The data stream and expert selection in WDMoE. When a user sends a prompt to an MoE-based LLM, it is preprocessed, embedded, and subjected to attention operations either locally or at BS, depending on user preference. Each token's embedding is then analyzed by a gating network to assign weights to each expert. WDMoE dynamically adjusts these weights and the number of experts based on gating network output and user channel conditions at the BS. The token embeddings are sent to the respective user devices for processing by expert networks. Once processed, the results are sent back to the BS, where they are weighted, summed, and then transformed from embeddings into words.}
    \label{figure_mechan}
    \end{figure*}

Retraining an LLM to account for wireless communication conditions poses significant challenges, particularly in ensuring the model can recognize channel conditions and adjust accordingly in distributed scenarios. Moreover, retraining the gating network of an MoE layer is not advisable due to its time-consuming and computationally intensive nature. Additionally, integrating wireless communications may inadvertently misguide the gating network, transforming the original training's single objective optimization problem into a multi-objective optimization problem, which can hinder model convergence.

Observing the sparsity of LLMs activations during inference and the occasional poor performance when transferring a trained model to out-of-distribution (OOD) datasets, we acknowledge that a decrease in the number of participating experts and a suboptimal selection of experts in the trained gating network can be tolerated. To address this dilemma, we design a training-free scheme that can modify expert selection without catastrophically impacting model performance, while also significantly reducing the latency.

The WDMoE mechanism is illustrated in Fig.\ref{figure_mechan}. A sequence of tokens enters the gating network, which then outputs weights for each expert in the MoE layer. Due to the frozen parameters of the router, the output weights are highly correlated with the model performance. The latency for devices can be estimated at BS, where BS acquires the channel conditions of all the devices and calculates the latency according to Eq.~\eqref{userl}. It is possible for some devices to be situated at a long distance from BS or in areas with significant coverage shadows. To address the potential issue of extended latency, we define a weight-to-latency ratio (WLR) for a token as follows:
\begin{equation}
    \mathbf{WLR}^i_{j} = \mathbf{w}^i_j \oslash \mathbf{t}^i_j,
\end{equation}
where $\mathbf{t}^i_j$ is the vector with elements $t^i_{j,q}$, for $q=1, 2, \cdots, n$. $\mathbf{WLR}^i_{j}$ is a vector consisting of $n$ experts, where $WLR^i_{j,q}$ denotes the WLR of the $q$-th expert. The operation $\oslash$ represents element-wise division. By jointly considering weights and latency, the WLR effectively balances them. Larger weights and lower latency contribute to higher WLR. While a large weight indicates a potential for better performance, excessive latency is intolerable for devices. Consequently, a mechanism based on WLR can filter out unsatisfactory expert networks based on both model performance and latency. The proposed scheme dynamically selects any number of experts as the processing node for the current token based on WLR.

Let $\theta$ be the threshold that determines whether to drop the expert with minimal WLR. Experts are sorted from the highest weights to the lowest weights, and the top $k$ experts are selected. Their WLRs are denoted as $\mathbf{WLR}^{i,k}_j$. Identify the minimal element $\min_{q\in \{1, 2, \cdots, k\}} \{WLR^{i,k}_{j,q}\}$, and the maximal element $\max_{q\in \{1, 2, \cdots, k\}} \{WLR^{i,k}_{j,q}\}$, within $\mathbf{WLR}^{i,k}_j$. These elements are used to calculate the decision variable $\kappa$ as follows:
\begin{equation}
    \kappa = \frac{\min\limits_{q\in \{1, 2, \cdots, k\}} \{WLR^{i,k}_{j,q}\}}{\min\limits_{q\in \{1, 2, \cdots, k\}} \{WLR^{i,k}_{j,q}\} + \max\limits_{q\in \{1, 2, \cdots, k\}} \{WLR^{i,k}_{j,q}\}}.
    \label{mecha}
\end{equation}
When $\kappa \leq \theta$, the expert with lowest WLR will be dropped.

\addtolength{\topmargin}{0.04in}
\begin{table*}[t]
    \centering
    \caption{Performance of different models on various benchmarks.}
    \label{performance}
    \begin{tabular}{lcccccccccc}
    \hline
    Model & Active Params & MMLU & PIQA & Arc-e & Arc-c & Humaneval & GSM-8K & BoolQ \\
    \hline
    Llama 2 7B & 7B & 46.8\% & 78.3\% & 56.1\% & 40.3\% & 12.8\% & 16.7\% & 74.9\% \\
    Llama 2 13B & 13B & 55\% & 79.8\% & 71.8\% & 60.3\% & 18.9\% & 29.6\% & 82.4\% \\
    Llama 2 70B & 70B & 69.7\% & 82.5\% & 85.9\% & 78.3\% & 26.2\% & 63.5\% & 87.7\% \\
    Mistral 7B-v0.1 & 7B & 64.1\% & 81.6\% & 83.6\% & 74.2\% & 22.6\% & 47.5\% & 84.1\% \\
    Mixtral 8x7B-Instruct-v0.1 & 13B & 70.0\% & 83.2\% & \textbf{92.8}\% & 84.8\% & 47.6\% & 70.9\% & \textbf{88.72}\% \\
    WDMoE-0.1 & 7/13B & \textbf{70.3}\% & \textbf{83.7}\% & 92.2\% & \textbf{88.1}\% & \textbf{48.8}\% & \textbf{71.0}\% & 88.38\%  \\
    WDMoE-0.2 & 7/13B & 68.8\% & \textbf{84.1}\% & 91.4\% & \textbf{86.8}\% & 47.0\% & \textbf{72.0}\% & 88.59\%  \\
    WDMoE-0.3 & 7/13B & 68.2\% & \textbf{83.7}\% & 88.9\% & 80.7\% & 45.7\% & 68.6\% & 87.06\%  \\
    \hline
    \end{tabular}
    \end{table*}

The threshold can be adjusted based on user requirements, with a higher threshold contributing to lower latency. It is important to note that a moderate increase in the threshold can reduce latency without a significant impact on model performance. However, excessively increasing the threshold can lead to a drastic drop in performance, despite achieving ultra-low latency. Additionally, this mechanism allows for the dynamic selection of any number of experts as needed and can simply repeat the process, as described in Algorithm \ref{alg:mechan}. Dropping an expert involves assigning zero weight to that expert, which means the corresponding device will not participate in processing.
\begin{algorithm}[t]
    \renewcommand{\algorithmicrequire}{\textbf{Input:}}
    \renewcommand{\algorithmicensure}{\textbf{Output:}}
    \caption{The WDMoE expert selection policy}
    \label{alg:mechan}
    \begin{algorithmic}[1]
    \REQUIRE $\mathbf{w}^i_j, \mathbf{t}^i_j, \theta$
    \ENSURE $\mathbf{w}^i_j$
    \STATE Calculate $WLR^i_{j,q} := \mathbf{w}^i_j \oslash \mathbf{t}^i_j$
    \STATE Rank experts in descending order of their weights
    \STATE Take the top $k$ experts' WLR as $\mathbf{WLR^{i,k}_j}$
    \STATE Calculate $\kappa$ according to Eq.~\eqref{mecha}
    \IF{$\kappa < \theta$}
    \STATE Drop the expert with minimal WLR
    \STATE Update the weights of experts
    \ENDIF
    \RETURN $\mathbf{w}^i_j := \text{Softmax}(\mathbf{w}^i_j)$
    \end{algorithmic}
    \end{algorithm}

\section{Experiment Results}
\subsection{Experiment Settings}
In the simulation, we consider an MEC-server deployed at BS and 8 devices. An expert network in each MoE block is deployed on one device, i.e. the expert network $q$ in each MoE layer is deployed on the $q$-th device. The distance between the $q$-th device and BS is denoted as $d_{q}$. We consider Rayleigh fading channels with a mean $10^{-\frac{PL(d_{q})}{20}}$, where the path loss is $PL(d_{q})(dB) = 32.4 + 20log_{10}(f^{carrier}) + 20log_{10}(d_{q})$. The carrier frequency $f^{carrier}$ is set as $3.5$ GHz. The transmission power of BS and the device is 10 Watts and 0.2 Watts respectively. The total bandwidth is 100MHz, allocated evenly among all devices. 

We apply the proposed WDMoE on Mixtral 8x7B model and refer to the modified version as \textbf{WDMoE-threshold} in the experiment results, with the specific numerical value for the threshold to be provided later. We leverage OpenCompass Platform\cite{2023opencompass} to conduct an in-depth and holistic evaluation of large language models. The benchmarks include MMLU\cite{hendrycks2020measuring}, PIQA\cite{bisk2020piqa}, ARC-Easy, ARC-Challenge\cite{clark2018think}, Humaneval\cite{chen2021evaluating}, GSM-8K\cite{cobbe2021training}, BoolQ\cite{clark2019boolq}. We implement the experiments on NVIDIA A40 GPUs.

\subsection{Performance Evaluation}
We compare WDMoE with current state-of-the-art models, including Llama 2 with 7B, 13B and 70B parameters, as well as Mistral (referred to as Mixtral 7B-v0.1) and Mixtral (referred to as Mixtral 8x7B-Instruct-v0.1), across various benchmarks. Besides, we evaluate the end-to-end latency of Mixtral and WDMoE in the wireless network.

\textbf{Performance.} The detailed results are presented in Table \ref{performance}. Due to the dynamics of WDMoE, the MoE's total number of active parameters is either 7B or 13B, which is lower than Llama 2 13B and Mixtral in some cases. We can see from Table I that WDMoE outperforms Llama 2 70B across all evaluated benchmarks while utilizing only around 20\% parameters and is generally superior to Mixtral. WDMoE-0.1 attains the peak performance on MMLU, Arc-c, and Humaneval benchmarks. WDMoE-0.2 excels in achieving the top scores on PIQA and GSM-8K benchmarks. WDMoE-0.3 demonstrates negligible performance deterioration on diverse benchmarks. Taking the Arc-c benchmark as an example, WDMoE demonstrates significantly higher accuracy compared to Llama 2 70B and Mixtral, with scores of 88.1\% vs. 78.3\% and vs. 84.8\%, respectively. 

\begin{figure*}[t]
    \centering
    \subfloat[PIQA accuracy.]{\includegraphics[scale=0.42]{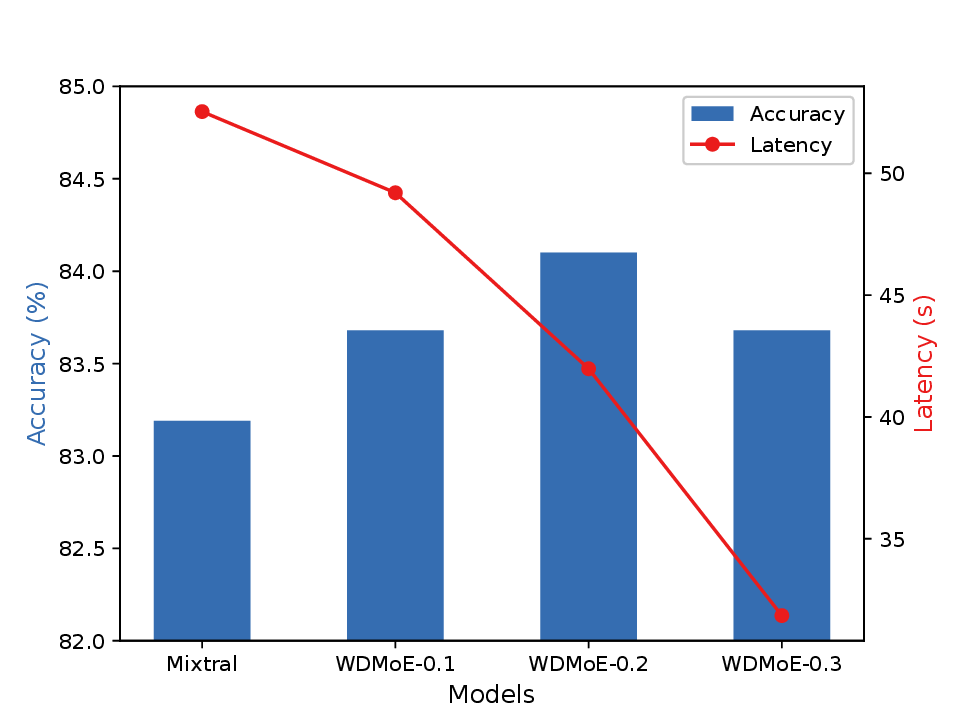}\label{piqa}}\hspace{5pt}
    \subfloat[ARC-C accuracy.]{\includegraphics[scale=0.42]{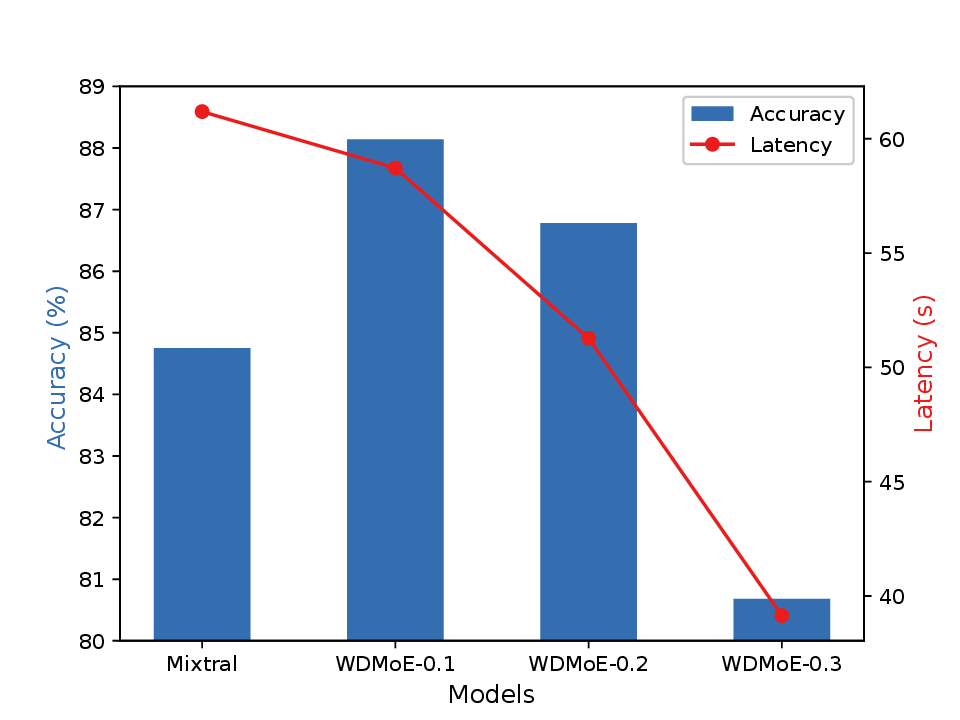}\label{arc-c}}\\
    \subfloat[HUMANEVAL accuracy.]{\includegraphics[scale=0.42]{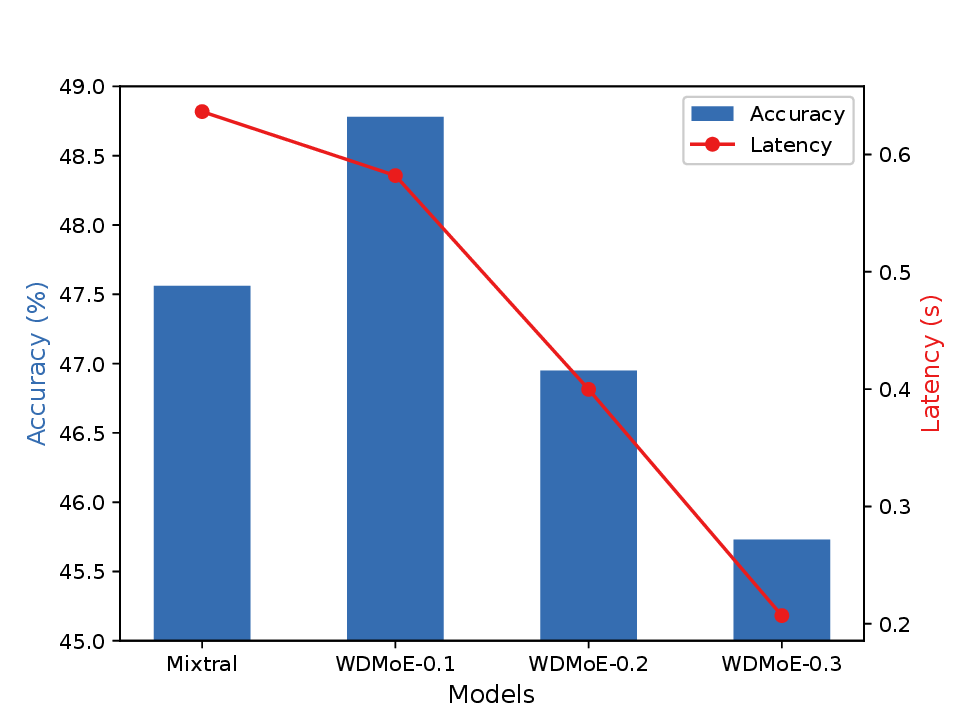}\label{humaneval}}\hspace{5pt}
    \subfloat[GSM-8K accuracy.]{\includegraphics[scale=0.42]{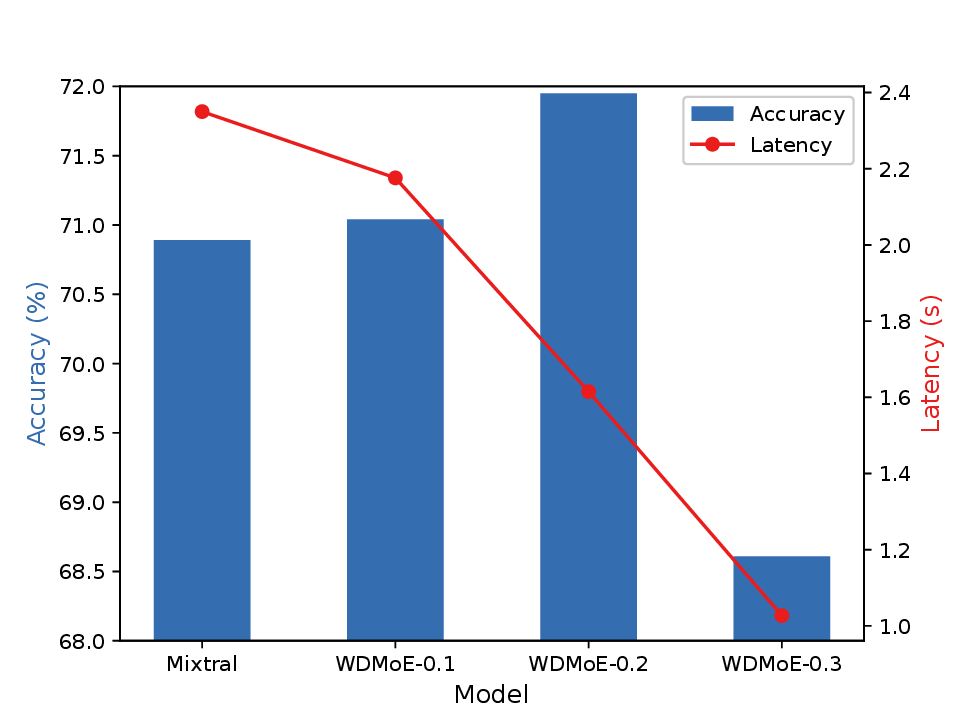}\label{gsm8k}}\\
    \caption{Performance and latency of WDMoE on various benchmarks.}
    \label{fig:results}
\end{figure*}

\textbf{Latency.} The end-to-end latency of Mixtral and WDMoE is evaluated in the wireless network. We deploy Mixtral and WDMoE on distributed devices, respectively and calculate the transmission latency and inference latency. The average end-to-end latency for one prompt, along with the corresponding model evaluation score results, is shown in Fig.\ref{fig:results}. The left axis represents model accuracy, and the right axis represents the end-to-end latency. The thresholds are set as 0.1, 0.2, and 0.3, respectively. As the threshold increases, the mechanism becomes more aggressive in dropping experts, and the latency is statistically lower. WDMoE-0.1 and WDMoE-0.2 have been demonstrated to outperform the current state-of-the-art on a majority of the evaluated benchmarks. WDMoE-0.2 guarantees 1.5\% gain in performance while achieving an average latency reduction of 30.21\% on GSM-8K benchmark. WDMoE-0.3 has acquired a higher performance score and is 1.65$\times$ faster than Mixtral on PIQA benchmark. 

Within a threshold range varying from 0 to 0.2, the model's performance remains relatively stable, and the latency is significantly reduced, by approximately 25.59\% on MMLU, 20.08\% on PIQA, 25.33\% on ARC-Easy, 16.2\% on ARC-Challenge, 37.54\% on Humaneval, 30.21\% on GSM-8K, and 21.86\% on BoolQ. When the threshold is increased to 0.3, the model's performance begins to deteriorate. Nevertheless, the slight decline in performance is offset by a notable reduction in latency, offering a more flexible choice for users. From the experiment results, we can conclude that WDMoE can significantly reduce the latency in wireless scenarios without severe performance deterioration.

\section{Conclusion}
In this paper, we propose a wireless distributed LLMs paradigm based on MoE, named WDMoE, which deploys LLMs collaboratively across edge servers at BS and mobile devices in the wireless communications system. The deployment framework leverages the parallel characteristics of expert networks. We also devise an expert selection policy that considers both the performance of the model and the end-to-end latency. Extensive experiments across various LLMs and on multiple benchmarks demonstrate that WDMoE ensures high performance and significantly reduces latency. This enables the feasibility of distributed LLMs in wireless scenarios and showcases the promising future of cooperative edge-device large models. 

\bibliographystyle{IEEEtran}
\bibliography{references}{}
\vspace{12pt}

\end{document}